\documentclass{mynature}
\usepackage{times}
\usepackage{graphicx}
\usepackage{amsmath}
\usepackage{amssymb}
\usepackage{textcomp} 

\begin{document}
\title{In-situ diagnostic of femtosecond probes for high resolution ultrafast imaging.}

\author{Chen Xie,$^{1,2\ast} $Remi Meyer,$^{2\ast}$ Luc Froehly,$^{2}$ Remo Giust,$^{2}$ Francois Courvoisier,$^{2\ast\ast}$\\
$^{1}$Ultrafast Laser Laboratory,\\ Key Laboratory of Opto-electronic Information Technology of Ministry of Education,\\
School of Precision Instruments and Opto-electronics Engineering,\\
Tianjin University, 300072 Tianjin, China\\
$^{2}$FEMTO-ST institute, Univ. Bourgogne Franche-Comt\'e, CNRS,\\
15B avenue des Montboucons, 25030, Besan\c{c}on Cedex, France\\
$^\ast$ These authors equally contributed\\
$^{\ast\ast}$ Corresponding author francois.courvoisier@femto-st.fr
}


\maketitle

\begin{abstract}

Ultrafast imaging is essential in physics and chemistry to investigate the femtosecond dynamics of nonuniform samples or of phenomena with strong spatial variations. It relies on observing the phenomena induced by an ultrashort laser pump pulse using an ultrashort probe pulse at a later time. Recent years have seen the emergence of very successful ultrafast imaging techniques of single non-reproducible events with extremely high frame rate, based on wavelength or spatial frequency encoding. However, further progress in ultrafast imaging towards high spatial resolution is hampered by the lack of characterization of weak probe beams. Because of the difference in group velocities between pump and probe in the bulk of the material, the determination of the absolute pump-probe delay depends on the sample position. In addition, pulse-front tilt is a widespread issue, unacceptable for ultrafast imaging, but which is conventionally very difficult to evaluate for the low-intensity probe pulses. Here we show that a pump-induced micro-grating generated from the electronic Kerr effect provides a detailed in-situ characterization of a weak probe pulse. It allows solving the two issues. Our approach is valid whatever the transparent medium, whatever the probe pulse polarization and wavelength. Because it is nondestructive and fast to implement, this in-situ probe diagnostic can be repeated to calibrate experimental conditions, particularly in the case where complex wavelength, spatial frequency or polarization encoding is used. We anticipate that this technique will enable previously inaccessible spatiotemporal imaging in all fields of ultrafast science and high field physics at the micro- and nanoscale.

\end{abstract}


\section{Introduction}

The fundamental understanding of laser matter interaction in several fields of ultrafast physics and chemistry requires imaging with both high spatial resolution (typ. sub 1~\textmu m), and high temporal resolution (typ. sub-100~fs). This is the case for instance for laser wakefield acceleration \cite{Matlis2006}, amplification in laser-excited dielectrics \cite{Winkler2017}, ultrafast ionization and plasma formation \cite{Hayasaki2017}, THz radiation \cite{Clerici2013,Zhang2018}, high harmonic generation \cite{Ghimire2018}, new material synthesis via laser-induced microexplosion \cite{Vailionis2011} or laser nanoscale processing \cite{Li2020,Pan2020}. 
The initial concepts of ultrafast imaging based on repetitive pump-probe measurements \cite{Keller1999,Krausz2001,Corkum2003,Zewail2007} have been recently complemented by a large number of different schemes allowing the imaging of non-reproducible events. This is performed via a sophisticated probe sequence or compressed photography \cite{Goda2009,Nakagawa2014,Gao2014,Li2014,Mikami2016,Ehn2017,Wang2020, Qi2020}, where the temporal information is encoded in the probe wavelength and/or in the spatial spectrum. 
 
However, further progress in ultrafast imaging at high resolution is still impeded by two problems. First, a key information is the absolute delay between pump and probe. This is crucial to link the excitation dynamics to the actual pump pulse intensity \cite{Krol2019}. Although synchronizing pump and probe pulses at a sample surface seems reasonably easy, the case of the synchronization of pump and probes in the bulk of a sample remains unaddressed. The problem is particularly acute when bulky microscope objectives impose pump and probe beams to pass through the same optic, in a nearly collinear geometry. In this case, a sample longitudinal shift by only 100 micrometers, such as the one needed to image through a 150~\textmu m microscope glass slide, shifts the relative delay between colinear 800~nm pump and 400~nm probe pulses by 40~fs because of the difference between their group velocities in the dielectric material. In other words, the absolute delay is intrinsically bound to the exact position of the focus inside the bulk of the solid or liquid medium under study. The pump-probe absolute delay must be determined using a pump-probe interaction on a scale of a few tens of micrometers to obtain a temporal accuracy in pump-probe measurement below the 10~fs scale. Unfortunately, conventional pulse synchronization techniques are inoperable in this context. Sum frequency generation and other nonlinear frequency mixing schemes require high intensities in the probe pulse. Frequency mixing can be operated only in specific crystals and with only a limited number of probe wavelengths. Polarization gating usually requires several 100's~\textmu m to several mm of pump-probe overlap distance for phase accumulation\cite{Kane+Trebino1993,Lebugle2015}. Transient-Grating cross-correlation frequency resolved optical gating (TG-XFROG) technique was successfully used to measure low energy probe pulses from a supercontinuum \cite{Lee+Trebino2008} or even in the UV \cite{Ermolov+Russell2016}. However, the conventional configuration is non-phase matched and the phase matching is only reached when the interacting waves have a sufficiently wide angular spectrum, {\it i.e} when they are focused, which is incompatible with the ultrafast imaging techniques mentioned above \cite{Goda2009,Nakagawa2014,Gao2014,Li2014,Mikami2016,Ehn2017,Wang2020, Qi2020}. Optical Kerr Effect cross-correlation was used in a spectral interferometry setup \cite{Sarpe2012} which is again incompatible with imaging.

A second particularly difficult issue in ultrafast imaging with high spatial resolution is the removal of pulse front tilt due to the angular dispersion of the probe pulse. This problem arises because of the necessary dispersion compensation of the temporal dispersion induced by microscope objectives \cite{Guild1997}. (As an example, a $\times$50 microscope objective induces a dispersion in excess of $\sim 12 000$~fs$^2$ at 400~nm, which stretches a 70~fs pulse to nearly 500~fs.) Most of the compensation schemes rely on spatially spreading the pulse spectrum such that unavoidable small misalignment creates angular dispersion \cite{Fork1984}. This can be usually neglected for very low numerical aperture. In contrast, in the case of high resolution imaging, the magnification of the setup also multiplies the coefficient of angular dispersion \cite{Chauhan+Trebino2010}. As we will see below, after a $\times$ 50 microscope objective, the pulse front tilt can easily exceed 70$^{\circ}$ (equivalent to a delay of 90~fs in 10~\textmu m field-of-view) for a misalignment of only 10 mrad in a prism compressor. Pulse front tilt can be measured using frequency conversion autocorrelation in various spatial schemes \cite{Sacks2001,Akturk2003,Grunwald2003,Bock2012} or using spatio-spectral interferometry \cite{Guang2014,Li2018} which are both unpractical for low-intensity pulses at arbitrary central wavelengths with $<$100~\textmu m beam diameter. Pulse front tilt is commonly regarded as very difficult to evaluate, particularly for low-intensity broadband pulses such as those used in the recent ultrafast photography techniques based on wavelength encoding.

 \begin{figure}[h!]
 \centering\includegraphics[width=\textwidth]{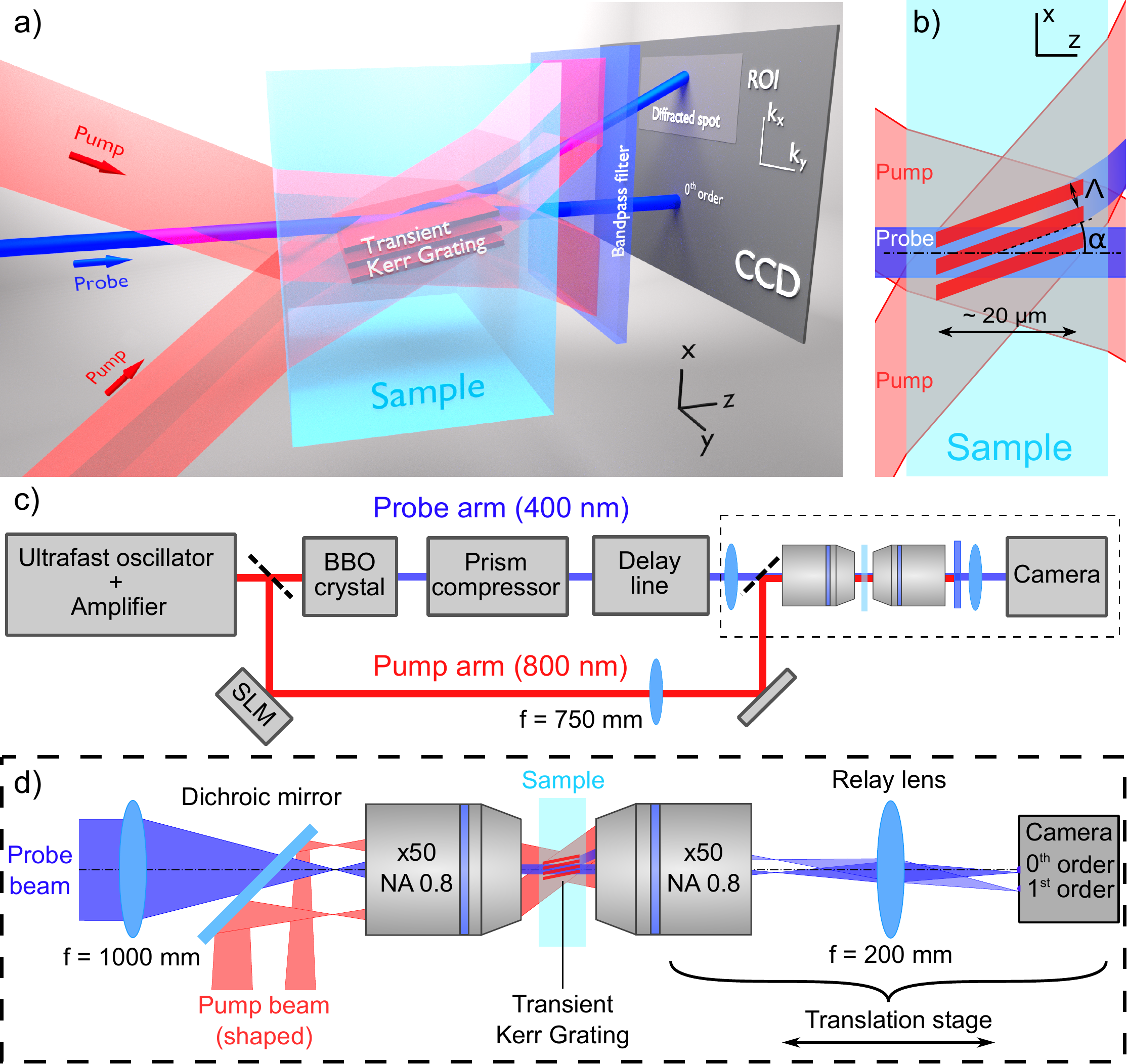}
 \caption{(a) Concept of the transient grating induced by the shaped infrared pump pulse in a transparent dielectric. The probe signal diffracted by the transient grating is collected in the far-field. (b) The Kerr-induced transient grating has a period $\Lambda$ and is tilted in respect to the probe axis by an angle $\alpha$. The length of the transient grating is a few tens of microns while the sample can be much thicker. (c) Experimental setup design. (d) Magnified view of the setup in the dashed box of (c) to show the interacting beams and the imaging configuration.}
 \label{fig:concept and setup}
 \end{figure}

Here we demonstrate a highly sensitive in-situ diagnostic for weak probe pulses which solves these two issues and is applicable to a large number of ultrafast imaging scenarios. The concept is shown in Figure \ref{fig:concept and setup}(a) and will be detailed in the next section. The pump pulse, shaped by a Spatial Light Modulator, creates a micrometric transient grating via the optical Kerr effect oriented in the Bragg condition for the probe beam. The micrometric size of the transient grating generates a pump-probe interaction that is highly localized which also allows for preserving a uniform diffraction efficiency over the broadband spectra of ultrashort probes. The diffracted signal provides a localized characterization of the absolute pump-probe delay. In addition, we have designed a way to straightforwardly visualize the angular dispersion using temporal stretching of the probe, so as to to efficiently remove pulse front tilt.
Our diagnostic is valid whatever the probe wavelength and polarization and uses intensities that are sufficiently low to avoid optical breakdown or damage. We have used probe pulse energy down to pJ level in the multishot integration regime. Therefore, this diagnostic can be repeated as many times as required, for instance at each sample replacement, to enable highly reproducible pump-probe imaging experiments.

The paper is organized as follows. We first derive the diffracted signal and present the optical setup. We then demonstrate that it can be used whatever the polarization configuration. We experimentally demonstrate we can retrieve the absolute pump-probe delay when the sample is longitudinally displaced and that the delay variation actually follows the difference of the group velocities between pump and probe. Last, we solve the second issue of pulse front tilt removal using a visualization tool based on pulse temporal stretching and observation of the diffracted signal in the far-field.

\section{Results}
\subsection{Cross-correlation signal and setup \label{sec:crossCorrSignal}}
We form a two-wave interference field inside a dielectric sample (fused silica, sapphire or glass) from a single pump beam, using a single Spatial Light Modulator, which automatically ensures the synchronization between the two pump waves. The instantaneous electronic Kerr effect transforms the interference intensity pattern into a grating with a period $\Lambda$ (see Fig. \ref{fig:concept and setup}(b)). We rotate the transient grating by an angle $\alpha$ to match the Bragg incidence condition for a probe pulse which is a collimated beam propagating along the optical axis. The rotation is simply performed by adding the same tilt angle $\alpha$ on the two interfering pump beams using the SLM. Typically, the crossing half-angle of the pump is $\theta =12^{\circ}$ and the rotation angle is $\alpha=6^{\circ}$. The ratio between these two angles is simply linked by the ratio between the probe and pump wavelengths (see Methods section). The interference pattern formed extends typically over a propagation distance of 40~\textmu m (see Suppl. Material Fig. \ref{fig:Beams}) and can be easily reduced by shrinking down the area of the SLM that directs light to the first order of diffraction. We can limit the interference pattern to below 10~\textmu m in length.

In the Methods section, we derive the diffraction efficiency, based on the coupled wave theory for thick gratings. The diffracted signal is a cross-correlation between the squared pump intensity and the probe intensity. The intensity in the first diffraction order, for a pump-probe delay $\tau$, reads :

\begin{equation}
    I_{1^{\mathrm{st}}} (\tau)  \propto  \bigg(\frac{ n_2 }{\cos \alpha}\bigg)^2\int  I_{\mathrm{pump}}^2(t) I_{\mathrm{probe}}(t-\tau) \mathrm{d}t \label{eq:CrossCorrationSingleEquation}
\end{equation}
\noindent with $n_2$ the Kerr index related to the relative polarizations states of pump and probe pulses.

 This signal will be key to characterize the probe, since the high-intensity pump pulse can be independently characterized with another technique. We note that the interaction is based on three plane waves, in contrast with conventional TG-XFROG where the phase matching is reached by crossing focused beams \cite{Lee+Trebino2008}. Preserving a probe beam as close as possible to a plane wave is important for further use in pump-probe imaging. The rotation of the interference field of the pump can be adapted to match the Bragg angle for any probe incidence, for instance to meet the requirements of ultrafast imaging with structured illumination \cite{Ehn2017}. In addition, this Bragg angle can be adjusted to any incident probe wavelength when encoding is based on the probe central wavelength in a very wide spectrum \cite{Goda2009}.

Our experimental setup is described in Figure \ref{fig:concept and setup}(c) and detailed in the Methods section. We use a Ti:Sapphire Chirped Pulse Amplifier (CPA) laser source which delivers $\sim 50$~fs pulses at 790~nm central wavelength and all measurements are performed by integrating the signal over 50 shots at 1 kHz repetition rate.

 We split the beam in a pump and a probe, which is frequency doubled with a $\beta$-Barium-Borate (BBO) crystal generating 60~fs pulses Full-Width at Half Maximum (FWHM). The pump pulse is then spectrally filtered to reduce its bandwidth to 12~nm FWHM, avoiding chromatic dispersion in the beam shaping stage. We spatially shape the pump beam using a Spatial Light Modulator (SLM) in near-normal incidence. The illuminated SLM is imaged by a 2f-2f telescope with a de-magnification factor of 208. The pump pulse duration has been evaluated to $\sim$115~fs at sample site, after the first microscope objective. This element is, in contrast, highly dispersive for the probe beam at 395~nm central wavelength (on the order of $\sim$ 6000~fs$^2$). We therefore compensate the linear dispersion on the 395~nm probe with a folded two-prisms compressor \cite{Fork1984}. The probe beam is also de-magnified by a factor of 278 so that the probe beam has a waist of 12~\textmu m (Rayleigh range of 1.1~mm) in the sample. The polarization state of the pump and probe pulses are independently controlled by the rotation of half-waveplates.

After interaction in the sample, we collect the pulses with a second $\times$50 (N.A. 0.8) microscope objective (MO). A relay lens images the Fourier plane of the second microscope objective onto a camera, which consequently records the far-field of the diffracted beams, which spatially separates the different orders of diffraction. Figure \ref{fig:Beams} in Supplementary Material shows the characterization of the generated pump interference field and of the unperturbed probe beam.

\begin{figure}[htb]
    \centering
    \includegraphics[width=0.9\textwidth]{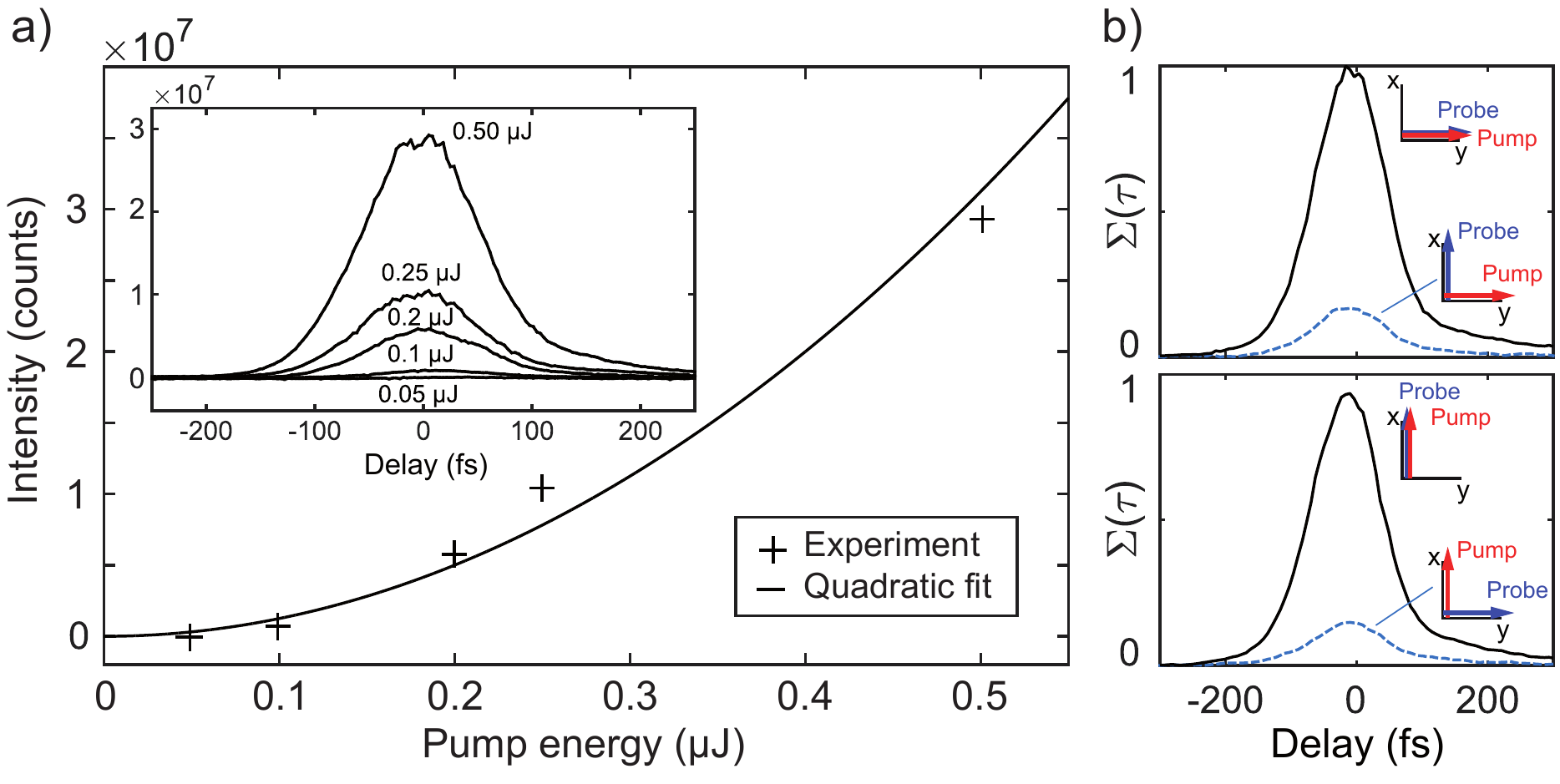}
    \caption{(a) Peak cross-correlation signal as a function of pump intensity. Crosses show experimental data and a quadratic fit is shown as dashed line. (inset) Cross-correlation signal as a function of pump-probe delay for different pump intensities, showing the peak position and shape are invariant with pump power. (b) Cross-correlation signal as a function of pump-probe delay for the 4 combinations of pump and probe polarization orientations. The parameters are provided in the Methods section.}
    \label{fig:Demonstration_Kerr}
\end{figure}

\subsection{A Kerr-based transient grating \label{sec:validation} valid for all combinations of pump-probe polarizations}

Here, we validate that the measured diffracted signal effectively follows Eq.\ref{eq:CrossCorrationSingleEquation} and demonstrate that the measurement is valid for all combinations of input pump and probe polarizations. We note that for sake of clarity, we report on the validation first, while in fact the experiments required to start with the removal of pulse front tilt followed by the optimization of the probe pulse duration, as will be explained below. Here, the probe pulse duration is 60~fs. In all figures below, the pump-probe delay is positive when the probe arrives after the pump pulse at the position of the transient grating in the sample. 

Figure \ref{fig:Demonstration_Kerr}(a) shows in the inset the recorded cross-correlation signal measured on the camera as a function of the relative pump-probe delay, for different pump pulse energies, indicated on top of each curve. Here, pump and probe pulses have the same horizontal polarization state. We observe that all curves have identical profiles, peaked at the same position. The main Figure \ref{fig:Demonstration_Kerr}(a) shows the evolution of the peak signal as a function of the pump pulse energy. It fits very well with a quadratic curve of the input pump energy as expected from Eq. \ref{eq:CrossCorrationSingleEquation}. The measurements have been performed in glass (Schott D263 microscope glass slide) and the results were also reproduced with identical conclusions in sapphire.
Therefore, at this input power level ($10^{11}$ to $10^{13}$~W$\cdot$cm$^{-2}$), no contribution from higher order nonlinearities or plasma formation is observable.

In Figure \ref{fig:Demonstration_Kerr}(b), we show the evolution of the cross-correlation signal for the four different linear polarization configurations: both pump and probe can be either horizontally or vertically polarized. The grating period is oriented vertically, as shown in Fig.\ref{fig:concept and setup}(a) . We checked that its orientation has no impact on the signal.

The effective Kerr index $n_2$ depends on the relative direction between pump and probe polarizations \cite{Loriot2009,Boyd2008}. Indeed, for an isotropic medium like glass, $n_{2{//}} = 3 n_{2{\perp}}$ where $n_{2 {//}}$ is the Kerr index when pump and probe polarizations are parallel, and $n_{2{\perp}}$ corresponds to the case where these polarizations are orthogonal to each other. Therefore, the signal efficiencies between parallel polarizations and orthogonal polarizations follow the ratio $\big(\frac{n_{2 //}}{n_{2\perp}}\big)^2 = {3^2}$. In our measurements, the signal ratio is in a range 6-10 when varying the grating period $\Lambda$. This ratio is highly sensitive to the background subtraction. Despite the relatively large error bar, the experimental ratio is in very good agreement with the expected one.
 
 These results overall confirm that the transient grating signal is effectively generated by Kerr effect. We therefore obtain the pump-probe synchronization using the barycenter of the curve. The cross-correlation curve also straightforwardly allows the measurement of the compression of the probe pulse while tuning the prism compressor. This is shown in Suppl. Fig. \ref{fig:PulseCompression}. It provides a direct evidence of the optimal compression for the probe at the sample site. In our case, the cross-correlation curve allows us to retrieve the probe pulse duration of $\simeq$60~fs FWHM knowing the pump pulse duration of 100~fs with 2300~fs$^2$ of second order dispersion (see Suppl. Mat.). We note that this in-situ diagnostic is also particularly useful when the sample itself is highly dispersive. 
 
 Finally, it is important to note that the technique is adaptable to characterize both polarizations. This is very useful to detect spectral phase differences in the optical path of the pump and probe beams. In Fig. \ref{fig:Demonstration_Kerr}(b), in all four polarization cases, the signal is effectively peaked at the same delay whatever the combination of input pulses polarizations. However, in preliminary experiments, a non-optimal dichroic filter used to recombine pump and probe had a different spectral reflectivity for vertical and horizontal pump polarizations, as shown in Supplementary Fig. \ref{fig:Dichroic}. For the horizontal pump polarization, it was inducing a spectral phase distortion. Our technique has identified this bias: a temporal shift as high as 100~fs and profile distortion was apparent from the cross-correlation curve. This highlights the effectiveness of the diagnostic even for the pump pulse.


\subsection{Spatial confinement of the synchronization}

Since pump and probe pulses usually have different group velocities in the sample, the synchronization criterion, {\it i.e.} the absolute zero pump-probe delay, has to be defined for a precise location of the focus in the sample.
In contrast with other synchronization or pulse measurement techniques, here, the interaction region between pump and probe is highly localized, down to some tens of micrometers. We successfully determined the pump-probe synchronization even for a transient grating length below 10 micrometers, obviously compromising on higher integration time to maintain an acceptable signal-to-noise ratio. The length of the transient grating can be adjusted using the SLM.

\begin{figure}[htb]
    \centering
    \includegraphics[width=0.8\textwidth]{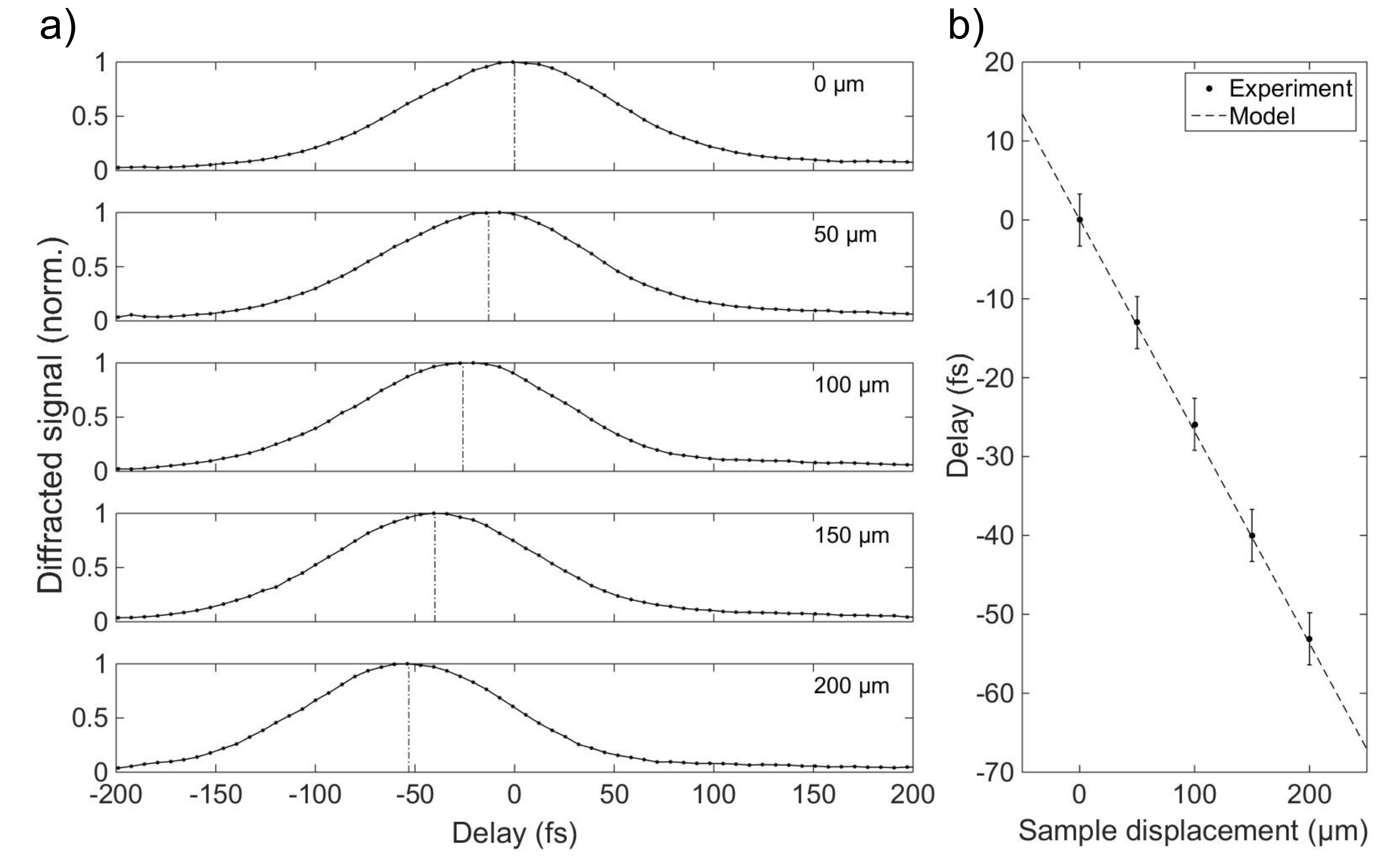}
    \caption{(a) Evolution of the TG signal as a function of sample position in sapphire (from 0 to 200 \textmu m). (b) Barycenter of TG signal as a function of sample displacement; experimental data are is excellent agreement with the model of Eq.(\ref{eq:delay}). The parameters are provided in the Methods section.}
    \label{fig:Depth}
\end{figure}

When we shift the position of the transient grating within the sample, we observe that the cross-correlation curve is shifted in delay. Experimentally, a first cross-correlation curve (Fig.\ref{fig:Depth}(a, top curve)) is acquired for a transient grating position starting at 50~\textmu m from the entrance surface of a 400~\textmu m thick sapphire sample, with refractive index $n_g^{790}=1.75$. The anisotropy of C-cut sapphire is negligible in comparison with the other effects. When the sample is then shifted upstream by a distance $d=50~\mu$m, the fringe pattern is shifted downstream by $d(n_g^{790}-1)$ = 37~\textmu m because of refraction. When we repeat the cross-correlation measurement for different positions of the transient grating, we observe the cross-correlation shift in delay by 13~fs every 50~\textmu m shift. This corresponds to the difference in group velocities between 790 and 395~nm wavelength. Analytically, the delay induced by the group velocity difference between red and blue pulses is (see Methods): 
\begin{equation}
    \Delta t = (n_g^{790}-n_g^{395})\frac{ d n_g^{790} }{c}
    \label{eq:delay}
\end{equation}
\noindent 
where $n_g^{395} = 1.796$ and $n_g^{790} = 1.750$ are the group indices of sapphire at the central wavelengths of 395 and 790~nm \cite{Querry1985}. We plot this curve in Fig. \ref{fig:Depth}(b) as a dashed line and see that it perfectly fits with the experimental data of the position of the barycenter of the cross-correlation curves reported from Fig. \ref{fig:Depth}(a). Similarly, we obtained an excellent agreement in Fused silica (see Suppl. Fig. \ref{fig:Depth_FS}), where the temporal delay is 22.6~fs every 100~\textmu m longitudinal shift. In microscope glass, the same shift induces a delay as high as 37~fs. We therefore demonstrate here that the strong localization of our measurement allows for retrieving the effect of the difference in group velocities on the pump-probe synchronization.

\subsection{Diagnostic for the pulse front tilt of the probe beam \label{sec:PFT}}

A prism compressor is aberration-free only when the two prisms are perfectly parallel. But it is experimentally unavoidable that the parallelism deviates by several milliradians. This deviation has however a dramatic impact on the probe pulse since it generates pulse front tilt, which is highly detrimental for the imaging of ultrafast phenomena. 
We will see here that the transient grating offers a straightforward visualization of the pulse front tilt such that it can be effectively canceled with the correct adjustment of the parallelism between the compressor prisms.

To evaluate how critical the problem is, we have evaluated the impact of a deviation angle from perfect parallelism between the two prisms in the prism compressor using {\sc Zemax}\textsuperscript{TM} software (see Methods section). After the prism compressor, for a deviation angle of 10~mrad of the second prism and compensation of the pointing direction with the folding mirror, the angular dispersion is 0.0045~mrad/nm with a negligible spatial chirp. However, the telescope used afterwards to decrease the probe beam waist to 12~\textmu m, increases the angular dispersion by the inverse of the magnification \cite{Chauhan+Trebino2010}, {\it i.e.} a factor of 278. Quantitatively, at the focus of the microscope objective, the angular dispersion becomes 12~mrad/nm. Overall, a positioning error of only 10~mrad generates a significant pulse front tilt as high as 78 degrees which would dramatically blur the dynamics of ultrafast phenomena imaged. In the following, we will see how the transient grating can be used to detect and remove this strong pulse front tilt.
\begin{figure}[htb]
    \centering
    \includegraphics[width=\textwidth]{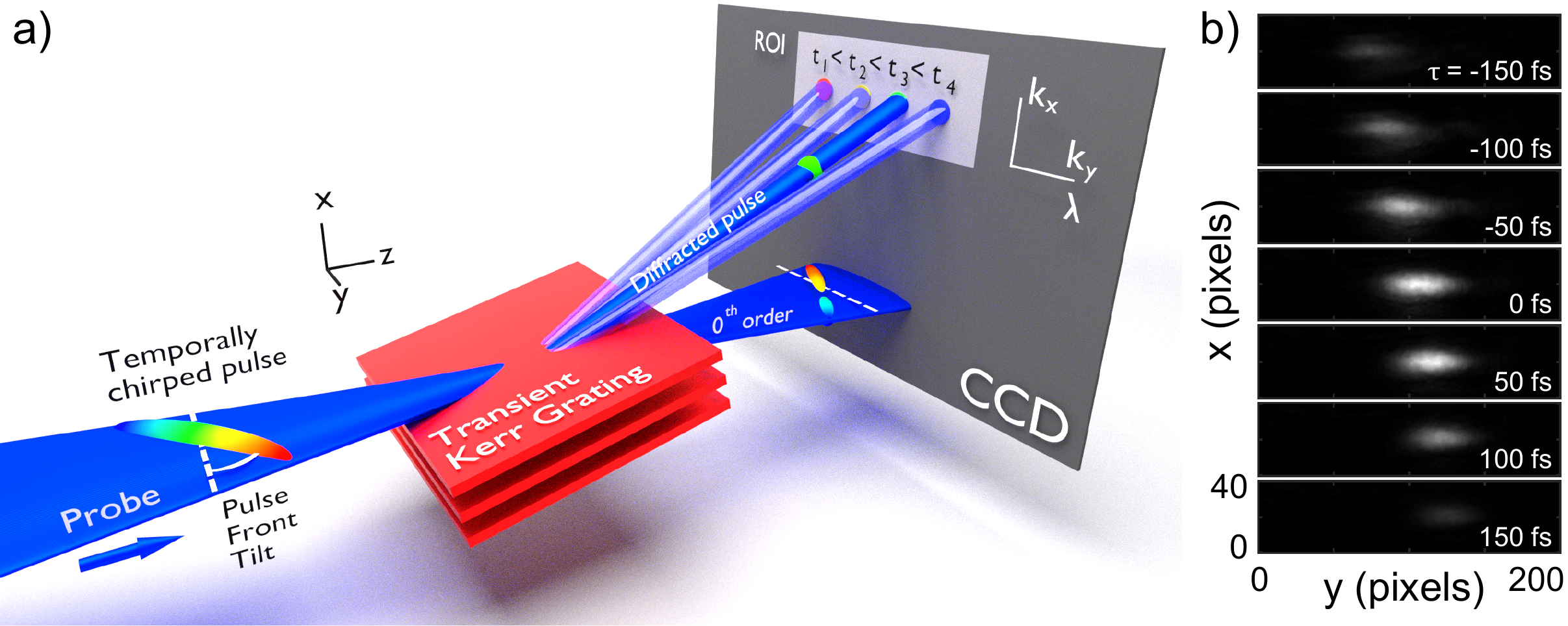}
    \caption{(a) Concept of the diffraction of an angularly dispersed probe pulse by the transient grating. The transient grating effectively samples the chirped pulse at the pump-probe delay and diffracts the corresponding sub-pulse on the ROI (Region of Interest) in the first order of diffraction. (b) Typical experimental result. Diffracted signal as a function of delay and deviation angle in $y$ direction.}
    \label{fig:AngularDispersion}
\end{figure}

To detect angular dispersion and remove pulse front tilt, we develop a technique based on the fact that angular dispersion acts as a spectrometer. It spreads the spectral content of the probe pulse on the horizontal $y$ axis, which corresponds to the direction of the angular mismatch in our prism compressor. Since the camera is placed in the Fourier space, each direction $k_y$ is mapped onto a single column of pixels. When the probe pulse is temporally chirped, the transient grating samples the probe pulse spectrum in time. This is sketched as a concept in Fig.\ref{fig:AngularDispersion}(a): the different wavelengths are sampled by the transient grating at different moments (because of temporal chirp) and are diffracted into different directions (because of angular dispersion). Figure \ref{fig:AngularDispersion}(b) shows a set of experimental images of the first diffracted order at different pump-probe delays when the probe pulse is slightly away from the optimal temporal compression. We observe the lateral shift of the diffracted spot along $y$ direction with the pump-probe delay, similarly as in the concept Fig. \ref{fig:AngularDispersion}(a). 
\begin{figure}[hbtp]
    \centering
    \includegraphics[width=\textwidth]{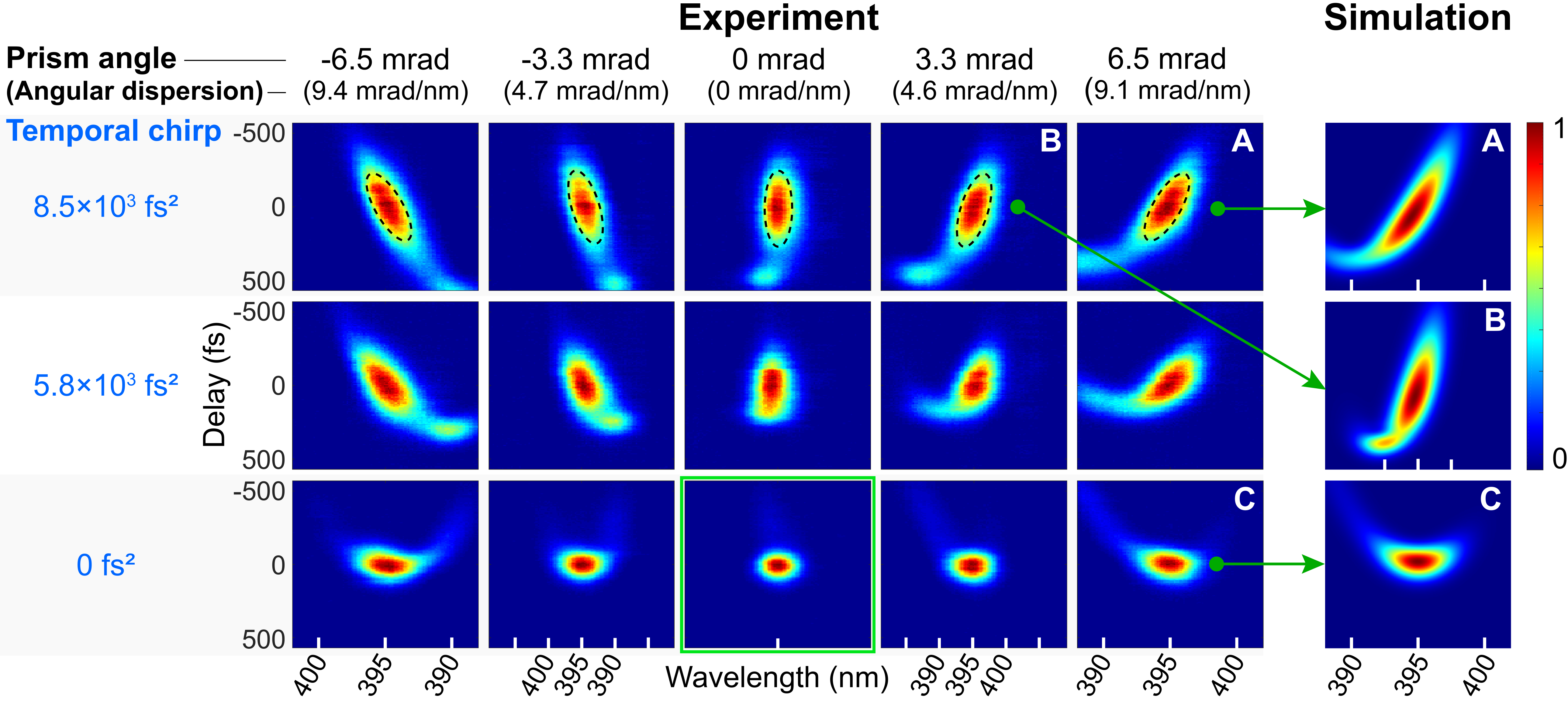}
    \caption{In the table, each trace shows the diffraction efficiency in arbitrary units as a function of delay (vertical axis) and spatial direction $k_y$ (horizontal axis, $k_y=[-1.03;1.03]$~\textmu m$^{-1}$). The left table shows experimental results for 15 different combinations of temporal chirp $\phi_2$ and angular dispersion. The angular dispersion has been numerically evaluated from the prism angle mismatch. The value of second order phase $\phi_2$ has been evaluated from the prism insertions in the prism compressor (first row 3~mm, second row 2~mm and last row 0~mm. The latter is the position for optimal pulse compression). For each trace, the horizontal axis scale has been converted to wavelength using the angular dispersion coefficient. When the angular dispersion is removed (central column), all wavelengths have the same direction $k_y$. In this case, the lateral width of the spot is simply determined by the Gaussian beam size. To show the consistency of the results, the rightmost column show three cases (A,B,C) where analytical formula for the diffraction efficiency of the transient grating has been integrated using the parameters extracted from the {\sc Zemax} simulations of the misaligned prism compressor (see Methods). }
    \label{fig:AngularDispPhi2}
\end{figure}
In Fig. \ref{fig:AngularDispPhi2}, we show the diffracted intensity as a function of $k_y$ (converted in wavelength by angular dispersion) and pump-probe delay for different values of angular dispersion (adjusted with prism angle) and temporal dispersion $\phi_2$ (adjusted with prism insertion). A single trace is obtained by summing for each delay the diffracted signal shown in \ref{fig:AngularDispersion}(b) along $x$-direction. Because of the angular dispersion acting as a spectrometer, these maps are similar to TG-XFROG signal maps, except this time in a phase-matched configuration because of the Bragg orientation.

When the spectral phase of the probe pulse is purely of second order, we have analytically demonstrated, in the Suppl. Material, that the signal in $(k_y,t)$ space appears as an ellipse. We have derived the slope of its major axis in the limit of high chirp and small angular deviation. The slope of the major axis of the ellipse expresses as (see Suppl. materials): 

\begin{equation}
    \frac{{\rm d} t}{{\rm d}k_y} =\frac{\phi_2}{p}
\end{equation}
\noindent where $\phi_2$ is the second order dispersion and $p$ the pulse front tilt \cite{Akturk2004}.

The top row of Fig. \ref{fig:AngularDispPhi2} shows the signal traces for a dispersion $\phi_2 \simeq 8.5 \times 10^3$~fs$^2$ at the sample site. The influence of the angular dispersion $p$, and the corresponding pulse front tilt, is highly apparent on the orientation of the major axis of the ellipse, which has been traced in dashed black line as a guide to the eye. The angular dispersion can be decreased to 0 with a sensitivity on the angular position of the prism as high as 0.5~mrad. The general symmetry observed between positive and negative values of angular dispersion is due to the mapping of the different wavelengths in increasing or decreasing order. We also note that a slight asymmetry in the value of angular dispersion between positive and negative values of prism angle is due to the spectral dispersion imposed by the first prism.

Therefore, our procedure to accurately visualize the angular dispersion is as follows. Since angular dispersion is proportional to the pulse front tilt $p$, we increase in a first step the temporal dispersion to make the inclination of the ellipse more apparent. We can then accurately remove the pulse front tilt. Finally, the prism insertion is adjusted to minimize the second order dispersion as shown for the signal framed in green ($\phi_2=0$, $p=0$).

In more detail, our numerical simulations show that the misalignment from perfect parallelism of the prisms not only affects the angular dispersion, but also introduces slight second and third order phases in the probe pulse. However, the variation of $\phi_2$ over the range of prism misalignments (a line in Fig.\ref{fig:AngularDispPhi2}) is typically of $\pm 1000$~fs$^2$, reasonably smaller than the values of second order dispersion that we introduced using the prism insertion. The third order phase is typically $10^5$~fs$^3$ and is apparent as a parabolic-shaped intensity pattern in the $(t,k_y)$ traces.

We have numerically simulated the experimental cases A,B,C of Fig.\ref{fig:AngularDispPhi2}. The analytical expression of the diffracted signal for a temporally chirped probe with higher-order phase is provided in Methods. We used for the probe's parameters the second and third order phases and the angular dispersion determined by the {\sc Zemax} modelling of the misaligned compressor. The results of our simulations are shown as the rightmost column in Fig.\ref{fig:AngularDispPhi2}. We find an excellent agreement between simulations and experiments.

We note that when the transient grating is rotated by 90 degrees, no variation in the arrival time is observable, since no element in our setup generates angular dispersion in this direction. Therefore, the transient grating diagnostic is particularly helpful to accurately remove pulse front tilt even for faint changes in the deviation angle of the prism compressor.


\section{Discussion}

We have developed an extremely localized in-situ diagnostic that allows a characterization and synchronization of a weak probe pulse with a higher intensity pump: the synchronization between pump and probe can be defined in a spatial domain of less than 10~\textmu m longitudinally inside the sample; we find the optimal point of probe pulse compression and the pulse front tilt can be removed. Therefore, this approach is extremely valuable for providing well-characterized ultrashort probe pulses for pump-probe imaging of ultrashort events under high magnification using single or multiple ultrashort probes over a wide spectrum and with different directions whatever the polarization. 

The use of an SLM is highly beneficial since the convenient switch between arbitrary phase profiles makes it possible to combine our diagnostic with pump-probe techniques where structured beam are involved. These structured beams indeed have a number of applications, such as high aspect ratio micro-nano structuring \cite{Gattass2008,Bellouard2004,Bhuyan2010,Courvoisier2016,Stoian2018}, laser welding \cite{Watanabe2006} among many others, and opens up new perspectives for studying laser-dielectric interaction in the ultrafast regime \cite{Hayasaki2012, Hayasaki2017,Bergner2018}. 

Our results have therefore a wide range of applicability, and we anticipate they will be particularly useful to characterize transient phenomena at micron-scale and laser-matter interaction within condensed matter.

\section*{Funding}
The authors acknowledge the financial supports of: European Research Council (ERC) 682032-PULSAR, Region Franche-Comte council (support to FRILIGHT platform), Labex ACTION ANR-11-LABX-0001-01; French RENATECH network and the EIPHI Graduate School ANR-17-EURE-0002.

\section*{Author contributions}
 R.M., R.G. and F.C. developed the setup on an initial concept by C.X and F.C., C. X. built the experimental interface for data acquisition, R.M. built the experimental setup, C.X. and F.C. performed the numerical simulations of beam propagation and diffracted signal.
 L.F. performed the numerical analysis of the pulse front tilt and dispersion using {\sc Zemax}. R.M., R.G. and F.C. analyzed the experimental data. F.C.made the analytical derivations. R.M. designed the figures. The manuscript was jointly written by C.X., R.M. and F.C. and revised by all authors.

\section*{Acknowledgments}
Technical assistance by L. Furfaro, C. Billet and B. Guichardaz as well as fruitful discussions with D. Brunner and J.M. Dudley from FEMTO-ST (Besan\c con, France) are gratefully acknowledged. 

\section*{Conflict of interest}
The authors declare no competing interests.

\section{Methods}

\textbf{SLM Phase Mask } The phase mask applied to our SLM is split in two equal parts: the top half generates a plane-wave-like beam propagating at an angle $\alpha - \theta$ and the lower half of the SLM mask symmetrically generates a beam propagating in direction $ \alpha + \theta$ toward the optical axis. The two beamlets cross at an angle of $2\theta$. We perform spatial filtering in the Fourier plane of the first lens ($f= 750$~mm) after the SLM to select only the first diffraction order due to the SLM mask. The fringe period of the interfering two pump waves is $\Lambda =\frac{\lambda_{pump}}{2 n \sin\theta}$. (For ease of reading, all angles are expressed in the dielectric medium of refractive index $n$, but wavelengths are expressed in vacuum)

\noindent \textbf{Bragg angle } The rotation angle $\alpha$ applied to match the Bragg incidence condition for the probe pulse, is determined by $ \lambda_{probe} = 2 n \Lambda \sin\alpha$, such that: $sin(\alpha)=sin(\theta) \frac{\lambda_{probe}}{\lambda_{pump}}$.

\noindent \textbf{Diffraction Efficiency }

The diffraction efficiency of the probe pulse on the pump-induced grating can be derived using coupled-wave theory describing thick gratings, since the grating is much longer than its period $\Lambda$ (see Suppl. Materials Fig.\ref{fig:Beams}). We use in the following the work by Kogelnik which provides analytically the effect of detuning \cite{Kogelnik1969}. We note that identical results could be obtained using Four-Wave Mixing (FWM) theory. The samples investigated here (fused silica and sapphire) possess a large bandgap. This ensures that resonant 3-photon absorption is negligible and therefore that the nonlinear Kerr response is instantaneous \cite{Steinmeyer2015}. We note that for TiO$_2$ and smaller bandgap dielectrics, the $\sim 6$~fs retardance observed in reference \cite{Steinmeyer2015} is close to negligible in comparison with the pulse durations used here.

The coupled wave theory allows for deriving the signal diffracted in the first order of diffraction. The assumptions are that the incident beam and diffracted one are monochromatic plane waves, incident on an infinitely wide grating of thickness $d$. Those conditions are reasonably fulfilled in our experiments. In this framework, the diffraction efficiency, {\it i.e.} the ratio between the diffracted intensity in the first order at the exit of the grating $|A_1(d)|^2$ and the incident beam intensity $|A_i|^2$, can be expressed as \cite{Kogelnik1969}:

 \begin{align}
    \eta(\xi, \phi) = \frac{\sin^2\sqrt{\xi^2+\phi^2}}{1+\bigg(\frac{\xi^2}{\phi^2}\bigg)}
    \label{eta}
\end{align}

\noindent where 
\begin{align}
     \phi = \frac{\pi}{\lambda(1-2\sin^2\alpha)^{\frac{1}{2}}}d \delta n
 \end{align}
 \begin{align}
     \xi = \frac{2\pi n \sin^2\alpha}{\lambda^2 (1-2\sin^2\alpha) } \delta\lambda
 \end{align}

\noindent $\lambda $ is the probe central wavelength. $\xi$ expresses the detuning, {\it i.e.} expresses how the diffraction efficiency reduces when the probe wavelength differs from the central wavelength at which the Bragg incidence is met. The wavelength detuning is written $\delta \lambda$. A similar relationship can also be expressed for the illumination angle detuning \cite{Kogelnik1969}. 

Therefore, in our experimental conditions, where $d \sim$ 30~\textmu m, $\delta n \simeq 10^{-4}$ and $\lambda = 0.4~\mu$m, $\alpha = 6^{\circ}$\, the diffraction efficiency is on the order of $10^{-3}$, which varies over a 30~fs probe pulse spectrum by less than $1\%$. (The diffraction efficiency peak width exceeds 50~nm FWHM, {\it i.e} about one order of magnitude larger than the bandwidth of a 30~fs pulse centered at 395~nm).

Since the wavelength detuning is negligible over the probe pulse spectrum, the diffraction efficiency becomes:

\begin{equation}
    \eta = \frac{|A_1(d)|^2}{|A_i|^2}=\sin^2\bigg( \pi \frac{d \delta n }{\lambda |\cos \alpha|}\bigg) 
\end{equation}

Then we can express the time-integrated diffracted intensity, with $\tau$ being the delay of the probe with regard to the pump pulse:
\begin{eqnarray}
    \Sigma (\tau) &=& \int I_{\mathrm{probe}} ^{\mathrm{1st~order}}(t) \mathrm{d}t \\
   &=& \int\sin^2\bigg( \pi \frac{d n_2 I_{\mathrm{pump}}(t)}{\lambda \cos \alpha}\bigg) I_{\mathrm{probe}}(t-\tau)\mathrm{d}t \\
    & \propto & \bigg(\frac{ n_2 }{\lambda \cos \alpha}\bigg)^2\int I_{\mathrm{pump}}^2(t) I_{\mathrm{probe}}(t-\tau) \mathrm{d}t \label{eq:CrossCorr}
\end{eqnarray}

\noindent provided that $d \delta n \ll \lambda \cos\alpha$ which is fulfilled in our experimental conditions. Hence, the diffracted signal is proportional to the correlation function between $I_{\mathrm{pump}}^2$ and $ I_{\mathrm{probe}}$, as in TG-XFROG \cite{Lee+Trebino2008, Sweetser+Trebino1997}.

\noindent \textbf{Setup} The Ti:Sapphire Chirped Pulse Amplifier (CPA) laser source (Coherent Legend) delivers $\sim 50$~fs pulses at 790~nm central wavelength and repetition rate 1~kHz. We split the beam in a pump and a probe, the latter is frequency doubled with a 50~\textmu m thick $\beta$ Barium Borate (BBO) crystal. The pump pulse is then spectrally filtered to reduce its bandwidth to 12~nm FWHM, avoiding chromatic dispersion in the beam shaping stage. The spectral transmission curve of the filter is nearly Gaussian so as to ensure the absence of pre-/post pulses. We spatially shape the pump beam using a Spatial Light Modulator (SLM) in near-normal incidence. The input beam is expanded to quasi-uniformly illuminate the full active area of the SLM. We de-magnify the resulting shaped beam by a factor 208 using a 2f-2f arrangement using a first lens of focal length $f_1 = 750$~mm and a second of focal length $f_2 = 3.6$~mm (Microscope Objective Olympus MPLFLN $\times$50 with Numerical Aperture (NA) 0.8). Spatial filtering of the first order of diffraction is performed in the Fourier plane of the first lens \cite{Froehly2014} (not shown in the figure). 

 We pre-compensate the linear dispersion of the 395~nm probe with a folded prism compressor \cite{Fork1984}. The probe beam is then de-magnified by a $\times$1/278 2f-2f arrangement (lens $f_3 = 1000$~mm and the same $\times$50 microscope objective) so that the probe beam in the sample has a waist of 12~\textmu m (Rayleigh range of 1.1~mm) at the focus of the microscope objective. 

Pump and probe energies and polarization states are controlled by half-waveplates and polarized beam splitters. Polarizations are controlled using independent half-waveplates placed on each beam path. The relative pump-probe delay is controlled with a motorized delay line in the probe beam, with a resolution of 3.3~fs. Figure \ref{fig:Beams} in the Supplementary Materials shows the pump and probe beam experimental characterizations.

After interaction in the sample, we collect the pulses with a second $\times$50 (N.A.~0.8) microscope objective (MO), as shown in Figure \ref{fig:concept and setup}(d). When recording the diffracted signal from the transient grating, we filter out the residual pump signal. A relay lens images the back focal plane of the second MO on a CCD camera, which consequently records the far-field of the diffracted beams, which spatially separates the different orders of diffraction. We select out the +1 order by summing the signal over a region of $\sim$100$\times$100 pixels on the CCD. The measurement is performed in multishot regime, with a 14 bits camera in free-run mode, with illumination time chosen in the range 10 to 50~ms so as to use the whole dynamical range of the camera over which the response is linear.   

 Before starting our experiments, we characterized the 790~nm pump pulse by self-referenced spectral interferometry (Wizzler\texttrademark) before the first microscope objective to measure and fully characterize the shortest pulse achievable, then we used the grating compressor of our CPA amplifier to compensate the dispersion of the MO and compress the pulse at the sample site, which was evaluated with two-photon generation in a BBO crystal. Taking into account the dispersion of the microscope objective of 2300~fs$^2$ at 790~nm central wavelength (separately characterized), the pump pulse duration is 100~fs at the sample site.
 
 \noindent \textbf{Shift in delay of the cross-correlation curve } When the sample is shifted upstream by $d$, the crossing point of the two pump beams is shifted downstream by a distance $\Delta = |d|(n_g^{790}-1)$, at first order in the small angle. Then, Eq.\ref{eq:delay} can be retrieved by considering the difference in optical paths between pump and probe from the initial case, again at first order in the small angles of the waves with the optical axis.
  
 \noindent \textbf{Prism deviation measurement } The relative angle of rotation of the second prism in the prism compressor has been monitored by measuring the deviation angle of the reflection of a laser pointer onto one facet of the prism. The precision of this measurement was 0.3~mrad.
 
  \noindent \textbf{Parameters for the experiments shown in the figures }
 \begin{table}[h]
     \centering
\begin{tabular}{|l|c|c|c|}
\hline 
 & Fig. \ref{fig:Demonstration_Kerr} & Fig. \ref{fig:AngularDispPhi2} & Fig. \ref{fig:Depth}  \\ 
\hline
Material & Schott D263 & Sapphire & Sapphire  \\ 
\hline 
$E_{pump}$ (\textmu J) & 0.05 - 0.5 & 0.1     & 0.1        \\ 
\hline 
$\theta$ ($^\circ$) & 14.4       & 12       & 12        \\ 
\hline 
\end{tabular} 
     \caption{Parameters used in the experiments shown in the figures above: material, pump energy $E_{pump}$ and half-crossing angle of the pump beams $\theta$. In all cases, to match the Bragg incidence angle, $\alpha$ is  half of $\theta$ because the probe central wavelength is half of the pump one.}
     \label{tab:table_param}
 \end{table}
 
 All angles are given in material and correspond to a single grating pitch $\Lambda=1.1$~\textmu m. Difference on angle values are related to respective material indices.

\noindent \textbf{Numerical simulations of the misaligned prism compressor }
  The prism compressor and the 2f-2f optical arrangement were numerically simulated using {\sc Zemax} software. This software enables a complete ray tracing over complex imaging systems and computes the optical path length. Hence it is then possible to model the dispersion effects up to the third order, including pulse front tilt and higher order dispersion. We note that the precise design of the Olympus MPLFLN $\times$50 microscope objective that we used experimentally is not available within this software. We replaced it by a $\times$60 microscope objective in our simulations. Since the dispersions are not precisely identical, we have adapted the dimensions of the folded prism compressor in the numerical simulations to precisely compensate the second order phase of the microscope objective and the lens. The parameters of the folded prism compressor were the following: the prisms are SF10 prisms with an apex angle of 60$^{\circ}$ degrees and a distance of 158 mm between the prisms. However, we note that the results obtained using this simulation are in excellent agreement with the experiments. The values of the higher order phases found using {\sc Zemax} and used to simulate the traces are respectively: (A) $\phi_2=1.1\times 10^4$~fs$^2$ and $\phi_3= -1.8\times 10^5$~fs$^3$, (B) $\phi_2=1.1\times 10^4$~fs$^2$ and $\phi_3=-1.6\times 10^5$~fs$^3$, (C) $\phi_2=3.1\times 10^3$~fs$^2$ and $\phi_3=-1.7\times 10^5$~fs$^3$.
  
 \noindent \textbf{Simulations of the diffracted signal in Fig. \ref{fig:AngularDispPhi2}} 
The simulations of the diffracted signal in Fig. \ref{fig:AngularDispPhi2} are based on numerical integration of the following expression :
$$ \Sigma (k_y, \tau)=\int | \hat{A}^{+1}(k_y,t) |^2 \mathrm{d}t \sim \iint |I_{\mathrm{pump}}.A_{\mathrm{probe}}(y,t)|^2 e^{i k_y y}\mathrm{d}t\mathrm{d}y$$
\noindent where $\hat{A}^{+1}$ is the spatial Fourier transform of the diffracted amplitude, and $A_{\mathrm{probe}}$ is the amplitude of the probe pulse.   $  \Tilde{A}_{\mathrm{probe}}(y,\omega)=E_0 e^{-T^2\omega^2/4} e^{-x^2/w_0^2} e^{-i p \omega y} e^{i (\frac{1}{2}\phi_2 \omega^2+\frac{1}{6}\phi_3 \omega^3+...)}$, following the model by Akturk {\it et al} \cite{Akturk2004}. 

The input pump pulse is modelled by a Gaussian pulse with 100~fs Full Width at Half-Maximum (FWHM). The probe is a 60~fs pulse with a small bump in the amplitude spectrum peaking at 390~nm so as to reproduce the experimental spectrum.

\bibliographystyle{naturemag}
\bibliography{TG_Biblio_abbrv}


\newpage
\setcounter{section}{0}
\bfseries
{\centering{{\Huge Supplementary Material}}}
\mdseries
\renewcommand{\thefigure}{S.\arabic{figure}}
\renewcommand{\thetable}{S.\arabic{table}}
\setcounter{page}{1}
\setcounter{figure}{0}
\setcounter{table}{0}

\section{Parameters for the experiments shown in the supplementary figures }

 \begin{table}[h]
     \centering
\begin{tabular}{|c|c|c|c|c|c|}
\hline 
&   Fig. \ref{fig:Beams} & Fig. \ref{fig:Dichroic} & Fig. \ref{fig:PulseCompression} & Fig. \ref{fig:Depth_FS}  \\ 
\hline
Material &   Air & FS  & Sapphire  & FS  \\ 
\hline 
$E_{pump}$ (\textmu J) &  $<$ 1 nJ    & 0.5          & 0.5 & 0.1  \\ 
\hline 
$\theta$ ($^\circ$) &   22       & 15              & 12 & 15 \\ 
\hline 
\end{tabular} 
     \caption{Parameters used in the experiments shown in the figures above: material, pump energy $E_{pump}$ and half-crossing angle of the pump beams $\theta$. In all cases, to match the Bragg incidence angle, $\alpha$ is  half of $\theta$ because the probe central wavelength is half of the pump one. FS stands for fused silica.}
     \label{tab:table_param_suppmat}
 \end{table}
 
 All angles are given in material and correspond to a single grating pitch $\Lambda=1.1$~\textmu m.

\section{Characterization of pump and probe beams}
\begin{figure}[htb]
    \centering
    \includegraphics[width  = 0.9\textwidth]{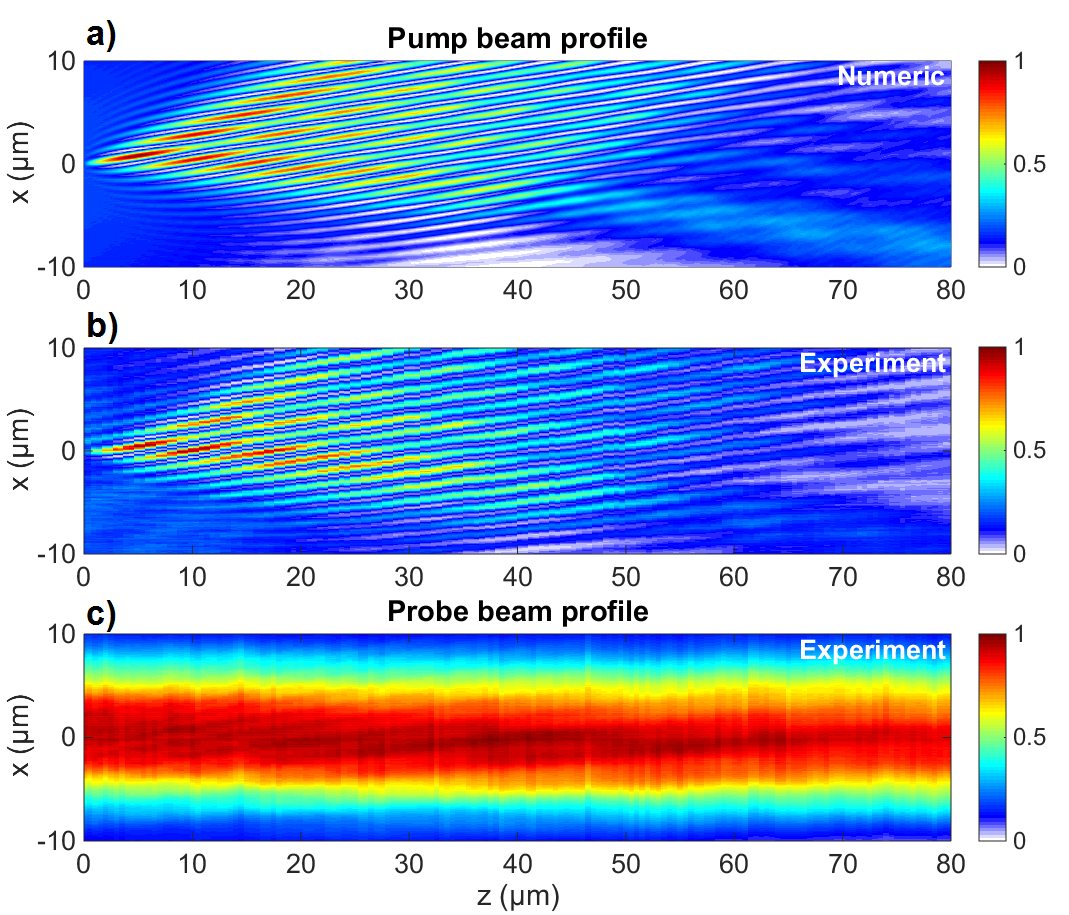}
    \caption{Fluence distribution in air of pump and probe beams along the propagation direction , {\it i.e.} in $(x,z)$ plane for $y=0$. (a) Numerical simulation for the pump beam using beam propagation method (b) Experimental characterization of the pump beam. (c) Experimental characterization of the unperturbed probe beam, in absence of pump.}
    \label{fig:Beams}
\end{figure}

Figure \ref{fig:Beams} shows the pump and probe beam characterizations in air. The pump beam is shaped by the SLM to generate an interference pattern as described in section \ref{sec:crossCorrSignal}, with $\theta = 22^{\circ}$ and $\alpha = 11^{\circ}$ in air. The grating period is therefore $\Lambda$ = 1.1~\textmu m. Experimental characterization and simulations of the spatial distribution are in excellent agreement.
 The probe beam is a homogeneous Gaussian beam. Our characterization shows that it is slightly wider than the fringe pattern and actually well superimposed over the fringes structure of the pump beam (in Fig. \ref{fig:Beams}, $x$ and $z$ scales are absolute distances, identical for the pump and probe beams).
 
 Technically, the experimental characterization of the near field is performed after replacing the last imaging lens of the setup described in Fig. \ref{fig:concept and setup}(d) by another convex lens with twice the focal length so as to image the near field of the focal plane of the microscope objective on the camera.

\clearpage
\section{Cross-correlation curve for a non-optimal dichroic filter}

  \begin{figure}[htb]
    \centering
    \includegraphics[width  = 0.75\textwidth]{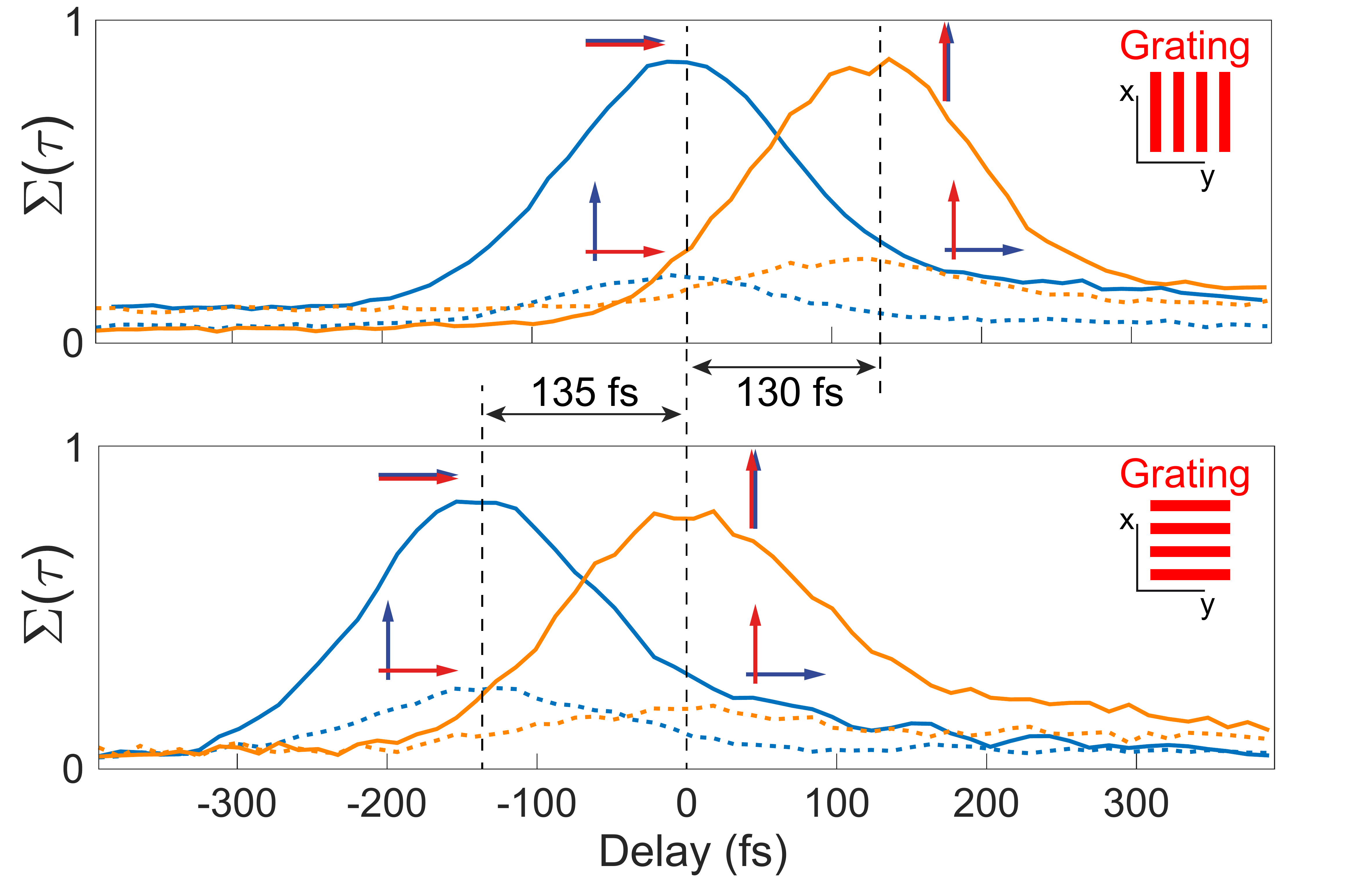}
    \caption{Evidence of a 133$\pm$3~fs shift induced by an imperfect dichroic mirror used for the pump-probe beams recombination. Blue and red arrows show respective polarization states of probe and pump beams.}
    \label{fig:Dichroic}
\end{figure}
This section shows a typical example where our technique shows particularly useful to identify potential flaws in the pump-probe timing. Here, the experiment is the same as described in the main text, except that the dichroic mirror shown in Fig.\ref{fig:concept and setup}(d) is a model Thorlabs DMSP650 instead of Layertech 101495. Figure \ref{fig:Dichroic} shows the diffracted intensity as a function of the pump-probe delay for all polarization combinations. This is presented, on top, for a configuration of the laser-induced grating where the periodicity is along $y$-direction. This is repeated in the bottom figure for a periodicity along $x$ direction, as in the results shown in the main text (see Fig.\ref{fig:Demonstration_Kerr}(b)).

The solid lines show the results for collinear pump and probe polarizations. Dashed lines are used for orthogonal polarizations.
We observe that in all cases, the signal for a horizontal pump polarization (blue curves) leads to a delay of approximately 130 fs with respect to the vertical one. This experiment shows that the dichroic mirror induces a distortion of the spectral phase for the vertical pump polarization important enough to make a retardance exceeding the pulse duration. In contrast, Fig.\ref{fig:Demonstration_Kerr}(b) in the main text shows the results in the same conditions for a different dichroic mirror (Layertech 101495) where the shift is absent.
We note that detecting this discrepancy between vertical and horizontal polarizations would be extremely difficult without our technique.

\section{Compression of the probe pulse}
 The cross-correlation signal provided by the transient grating allows for controlling the probe compression and estimating the probe pulse duration.

\begin{figure}[htb]
    \centering
    \includegraphics[width=0.8\textwidth]{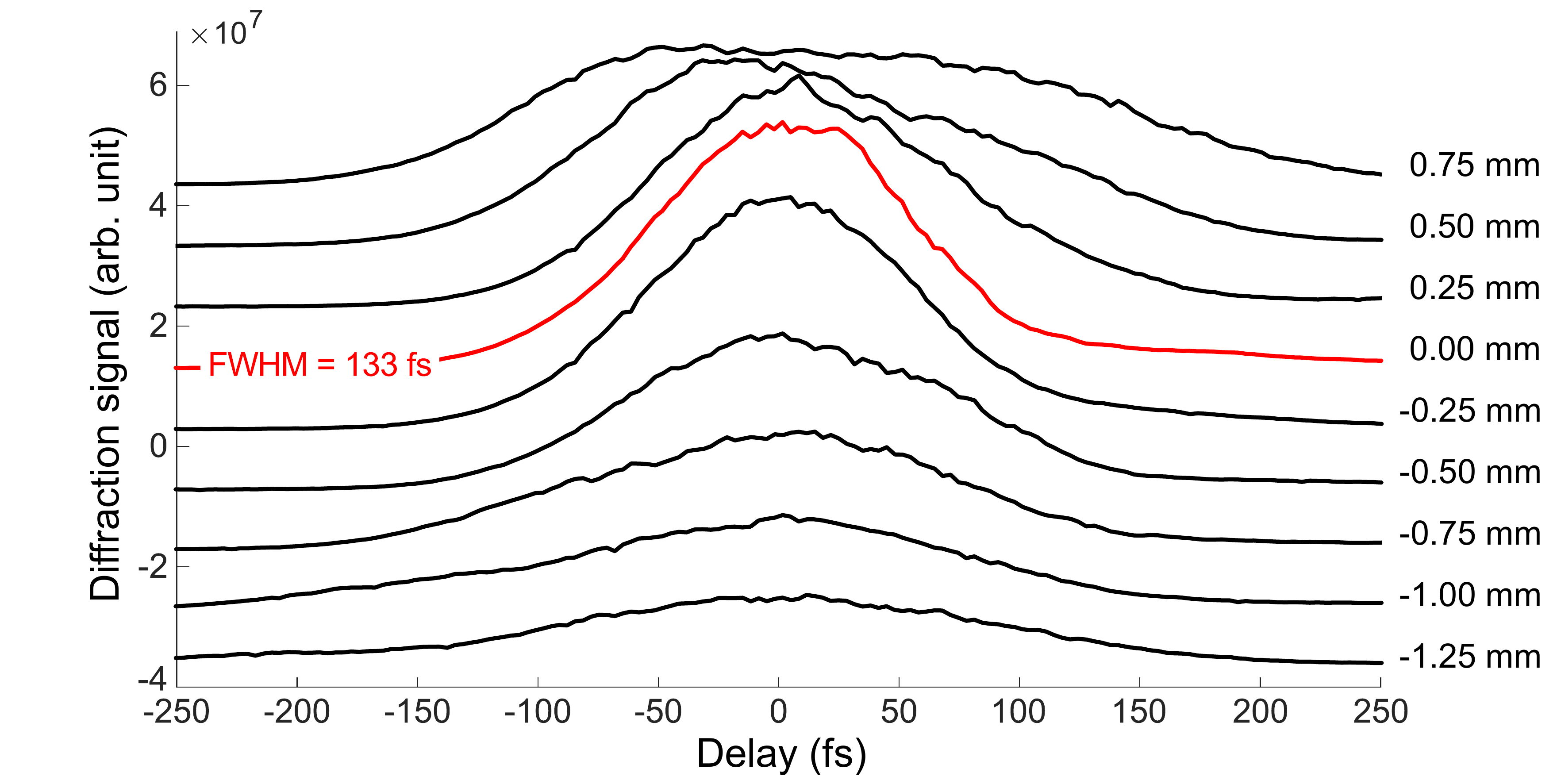}
    \caption{Signal for different positions of the prism insertion.}
    \label{fig:PulseCompression}
\end{figure}

In figure \ref{fig:PulseCompression}, we plot the diffracted signal as a function of delay for different positions of the prism compressor (insertion of glass in the probe beam path). We observe that for the position noted as ``0 mm", the cross-correlation signal is the most compressed. This corresponds to the shortest pulse duration achievable for the probe pulse.
Using the knowledge of the pump pulse duration ($\simeq$ 110~fs), we retrieve the probe pulse duration by numerical fitting of the experimental curve, which leads to a probe pulse duration of $\simeq$ 60~fs. 

\section{Synchronization in fused silica}

\begin{figure}
    \centering
    \includegraphics[width=0.9\textwidth]{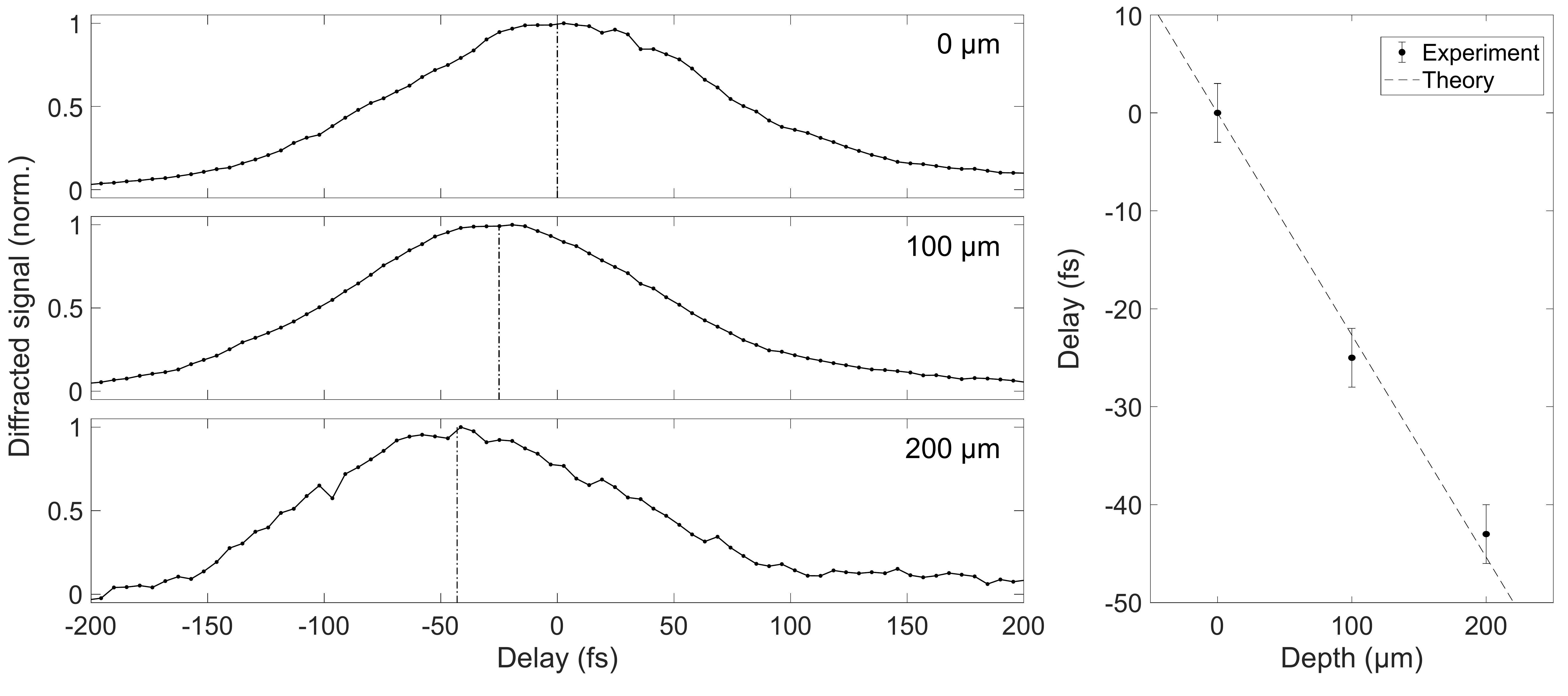}
    \caption{(left) Evolution of the diffracted signal as a function of sample position in fused silica (from 0 to 200 \textmu m). (right) Barycenter of the diffracted signal as a function of sample displacement; }
    \label{fig:Depth_FS}
\end{figure}

Figure \ref{fig:Depth_FS} shows, as in the main text Fig \ref{fig:Depth}, the evolution of the diffracted signal as a function of the pump-probe delay for three different positions of the sample. This time, it is performed for fused silica material instead of sapphire. We obtain similarly an excellent agreement between experimental data and the prediction of Eq.\ref{eq:delay}, {\it i.e.} -22.6~fs for 100 \textmu m.

\section{Inclination of the $t$-$k_y$ traces of Fig. \ref{fig:AngularDispPhi2} in presence of angular dispersion}

In this section, we analytically derive the inclination angle of the traces of Fig. \ref{fig:AngularDispersion}. 
We start with the expression of a pulse with a Gaussian distribution in space and time, with waist $w_0$ and with pure angular dispersion with parameter $p=\frac{\mathrm{d}k_y}{\mathrm{d}\omega}$\cite{Akturk2004} in the direction $y$ and second order phase $\phi_2$. The pulse duration (FWHM) is $\sqrt{2 log(2)} T$:
\begin{equation}
    \Tilde{E}(y,\omega)=E_0 e^{-T^2\omega^2/4} e^{-x^2/w_0^2} e^{-i p \omega y} e^{i \phi_2/2 \omega^2}
\end{equation}
\noindent In this equation, $t$ corresponds to the pump probe delay.

\noindent After double Fourier transformation over $y$ and $t$ coordinates:
\begin{equation}
    \hat{E}(k_y,t) =E_0' e^{-\alpha t^2 -\frac{(k_y -2 i \alpha p  t)^2}{4(1/w_0^2+p^2\alpha)}}=E_0' e^{u+i v}
\end{equation}

\noindent with $E_0'$ being a constant, $u$ and $v$ real-valued, and $$\alpha =\frac{1}{T^2 +2i\phi_2}=\alpha'+i\alpha''$$

\noindent 
We isolate $u$:

\begin{equation}
\begin{split}
    u=-\alpha' t^2 - & \frac{1}{\Delta}\Big[k_y^2-4 (p t)^2 (\alpha'^2-\alpha''^2)+4 k_y p t \alpha'' \Big] \times \Big[\frac{1}{w_0^2}+p^2 \alpha'\Big] \\
     + & \frac{(p)^2\alpha''}{\Delta} \Big[-4 k_y p t \alpha'-8 \alpha' \alpha'' p t\Big]
    \end{split}
\end{equation}

\noindent with $\Delta = 4 \Big[ \Big(\frac{1}{w_0^2} +p^2 \alpha' \Big)^2 +\Big(p^2 \alpha''  \Big)^2 \Big]$.

The location of the iso-intensity patterns detected on the camera is determined by $u=K$, where $K$ is a constant. This equation can be rewritten in a quadratic form, which is the equation of an ellipse. We introduce the normalized transverse wavevector $\tilde{k_y}$ and time $\tilde{t}$ with $k_y=\kappa \tilde{k_y}$ and $t=\tau \tilde{t}$.

\begin{equation}
\label{eq:elliptic}
    A \tilde{k_y}^2 + B\tilde{k_y}\tilde{t} +C \tilde{t}^2 +D\tilde{k_y}+ E\tilde{t}+F =0
\end{equation}

\noindent with the following values:
\begin{equation}
    \begin{split}
    A = & -\frac{\kappa^2 \xi}{\Delta} \\
    B = & \kappa\tau \frac{(4p\alpha''\xi-4p^3\alpha'\alpha'')}{\Delta} \\
    C = & \tau^2\frac{(\Delta \alpha' +4 p^2 \alpha''^2\xi-4 p^2\alpha'^2\xi -8\alpha'\alpha''^2p^4)}{\Delta} \\
    D = & 0\\
    E = & 0 \\
    F = & 0 
    \end{split}
\end{equation}
\noindent $F$ is chosen here to zero but it can be any constant and $\xi=\Big( \frac{1}{w_0^2} +p^2\alpha' \Big)$.

\noindent Equation \ref{eq:elliptic} describes an ellipse in $(k_y,t)$ space, which can be rewritten in a matrix form as:

\begin{equation}
    ^{T}X \mathbb{A}X + ^{T} \mathbb{B}X+F=0 
\end{equation}
\noindent with
\begin{equation*}
    \begin{split}
       X = &
\begin{pmatrix}
k_y \\
t
\end{pmatrix}\\
 \mathbb{A} = &
 \begin{pmatrix}
 A & B/2 \\
 B/2 & C
 \end{pmatrix}
 \\
 \mathbb{B} = &
 \begin{pmatrix}
 D \\
 E
 \end{pmatrix}\\
    \end{split}
\end{equation*}

The major axis of the ellipse in $(k_y,t)$ space is rotated by an angle $\theta$ from the $k_y$ axis. What follows is the determination of this angle.
The rotation matrix $R= \big(\begin{smallmatrix} \cos\theta & -\sin\theta\\ \sin\theta & \cos\theta \end{smallmatrix}\big)$ is chosen so that $^{T}R \mathbb{A} R$ is diagonal. 
The two eigenvalues of $\mathbb{A}$ are:
\begin{equation}
    \lambda_{1,2}=\frac{A+C\pm \sqrt{A^2+B^2+C^2-2AC}}{2}
\end{equation}

\noindent The eigenvectors allow the construction of the rotation matrix $R$. Then the rotation angle can be determined from:
\begin{equation}
\label{eq:tangent}
    \tan\theta=\frac{-B/2}{C-\lambda_2}=\frac{B}{A-C+\sqrt{A^2+B^2+C^2-2AC}}
\end{equation}

In the following, we perform a development of $A$,$B$ and $C$ assuming that the temporal dispersion $\phi_2$ is large and the angular dispersion $p$ is small. We use the following adimensional parameter: 
$$ \varepsilon = \frac{p^2T^2w_0^2}{\phi_2^2}$$

After a lengthy but straightforward calculation, we get at the first order in $\varepsilon$:

\begin{equation}
    \begin{split}
    A \simeq & \frac{\kappa^2 w_0^2}{4} \Big(-1+\frac{\varepsilon}{4}\Big) \\
    B \simeq & \kappa\tau \Big(1-\frac{\varepsilon}{4}\Big)\frac{p w_0^2}{2\phi_2}\\
    C \simeq &\frac{\tau^2}{4 T^2}\Big[-\frac{T^4}{\phi_2^2}+\varepsilon \Big(-1+\frac{5\tau^4}{4\phi_2^2}\big)\Big] \\
    \end{split}
\end{equation}

\noindent We then find the denominator of Eq. \ref{eq:tangent}:  
\begin{equation}
    A-C+\sqrt{A^2+B^2+C^2-2AC} \simeq \frac{\tau^4 T^4}{8\kappa^2w_0^2\phi_2^4} +\varepsilon \Big(\frac{\tau^2}{2T^2}-\frac{\tau^2T^2}{8 \phi2^2}+\frac{\tau^4}{4\kappa w_0^2 \phi_2^2}\Big) 
\end{equation}

\noindent This can be further simplified using the fact that $\phi_2\gg T^2$ such that
\begin{equation}
    A-C+\sqrt{A^2+B^2+C^2-2AC} \simeq  \frac{\tau^2}{2T^2} \varepsilon 
\end{equation}

\noindent We finally get:
\begin{equation}
    \tan\theta \simeq \frac{\kappa}{\tau}\frac{\phi_2}{p}
\end{equation}

\end{document}